# Helium in the eroding atmosphere of an exoplanet

J. J. Spake[1], D. K. Sing[1,2], T. M. Evans[1], A. Oklopčić[3], V. Bourrier[4], L. Kreidberg[5,6], B. V. Rackham[7,8,9], J. Irwin[6], D. Ehrenreich[4], A. Wyttenbach[4], H. R. Wakeford[11], Y. Zhou[7], K. L. Chubb[10], N. Nikolov[1], J. Goyal[1], G. W. Henry[12], M. H. Williamson[12], S. Blumenthal[1], D. Anderson[13], C. Hellier[13], D. Charbonneau[6], S. Udry[4], and N. Madhusudhan[14]

[1]*Astrophysics Group, School of Physics, University of Exeter, Stocker Road, Exeter, EX4 4QL, UK.*
[2]*Department of Earth and Planetary Sciences, Johns Hopkins University, Baltimore, MD, USA*
[3]*Institute for Theory and Computation, Harvard-Smithsonian Center for Astrophysics 60 Garden Street, MS-51, Cambridge, Massachusetts 02138, USA*
[4]*Observatoire de l'Université de Genève, 51 chemin des Maillettes, 1290 Sauverny, Switzerland*
[5]*Harvard Society of Fellows 78 Mt. Auburn St. Cambridge, MA 02138, USA*
[6]*Harvard-Smithsonian Center for Astrophysics 60 Garden St. Cambridge, MA 02138*
[7]*Department of Astronomy/Steward Observatory, The University of Arizona, 933 N. Cherry Avenue, Tucson, AZ 85721, USA*
[8]*National Science Foundation Graduate Research Fellow*
[9]*Earths in Other Solar Systems Team, NASA Nexus for Exoplanet System Science*
[10]*Department of Physics and Astronomy, University College London, London, WC1E 6BT, UK*
[11]*Space Telescope Science Institute, 3700 San Martin Drive, Baltimore, MD 21218, USA*
[12]*Center of Excellence in Information Systems, Tennessee State University,, Nashville, TN 37209, USA*
[13]*Astrophysics Group, Keele University, Staffordshire, ST5 5BG, UK*
[14]*Institute of Astronomy, University of Cambridge, Madingley Road, Cambridge CB3 0HA, UK*

**Helium is the second most abundant element in the universe after hydrogen and is a major constituent of gas-giant planets in our Solar System. Early theoretical models predicted helium to be among the most readily-detectable species in the atmospheres of exoplanets, especially in extended and escaping atmospheres[1]. However, searches for helium have until now been unsuccessful[2]. Here we present the first detection of helium on an exoplanet, at a confidence level of 4.5σ. We measured the near-infrared transmission spectrum of the warm gas giant WASP-107b[3] with the Hubble Space Telescope and identified the narrow absorption feature of excited, metastable helium at 10,833 angstroms. The amplitude of the feature, in transit depth, is 0.049±0.011% in a bandpass of 98 angstroms, which is more than 5 times greater than what could be caused by nominal stellar chromospheric activity. The large absorption signal suggests that WASP-107b has an extended atmosphere that is eroding at a total rate of $10^{10}$-$3\times10^{11}$ g s$^{-1}$ (0.1-4% of its total mass per Gyr), and may have a comet-like tail of gas shaped by radiation pressure.**

WASP-107b is one of the lowest density planets known, with a radius similar to that of Jupiter (0.94±0.02$R_J$) and a much lower mass (0.12±0.01$M_J$)[3]. It orbits an active K6 dwarf every 5.7 days at a distance of 0.055±0.001 astronomical units. On 31 May 2017, we observed a primary transit of WASP-107b with the Wide Field Camera 3 (WFC3) on board the Hubble Space Telescope (HST). Our observations lasted 7 hours and we acquired 84 time-series spectra with the G102 grism, which covers the 8,000 – 11,000 Å wavelength range. Further details of the observations and data reduction can be found in Methods. Each spectrum was integrated along the wavelength axis to first produce a 'white' light curve (Extended Data Fig. 1). In addition to the planetary transit signal, the resulting time series was affected by instrumental systematics caused by electron trapping in the WFC3 detector. We fitted the white light curve with a planetary transit model[14] multiplied by a linear baseline trend and a physically-motivated WFC3 systematics model[15]. For the planetary transit model, we allowed the planet-to-star radius ratio ($R_p/R_s$) and the mid-transit time ($T_0$) to vary as free parameters, while holding the ratio of orbital distance to stellar radius ($a/R_s$), inclination ($i$), eccentricity ($e$), and period ($P$), fixed to previously determined values[6,16]. We assumed a quadratic limb-darkening profile for the star, holding the coefficients fixed to values determined from a model stellar spectrum[17]. Further details of this fit are provided in

Methods. The results of the fit are reported in Extended Data Table 1, and Extended Data Fig. 1.

Two sets of spectroscopic light curves were constructed by summing each spectrum into broad- and narrow-band bins. The first set consisted of 9 broad-band channels spanning the 8,770-11,360 Å wavelength range, while the second set comprised 20 overlapping, narrow-band channels spanning the 10,580-11,070 Å wavelength range. The narrow-band channels cover the helium absorption triplet at 10,833 Å (vacuum wavelength – the air wavelength of this line is 10,830 Å). The widths of the broadband and narrowband channels were 294 Å (12 pixel columns) and 98 Å (4 pixel columns), respectively. We fitted both sets of spectroscopic light curves using the same approach as described above for the white light curve. However, for the planetary transit signals, we only allowed $R_p/R_s$ to vary as a free parameter, while holding $t_0$, $a/R_s$, $i$, $e$, and $P$ fixed to those reported in Extended Data Table 1. We fixed limb darkening coefficients in a similar way to the white light curve fit. Additional details of the fitting procedure are given in Methods. The inferred values for the transit depth, $(R_p/R_s)^2$, in each wavelength channel are shown in Fig. 1 and Extended Data Table 2. These results constitute the atmospheric transmission spectrum.

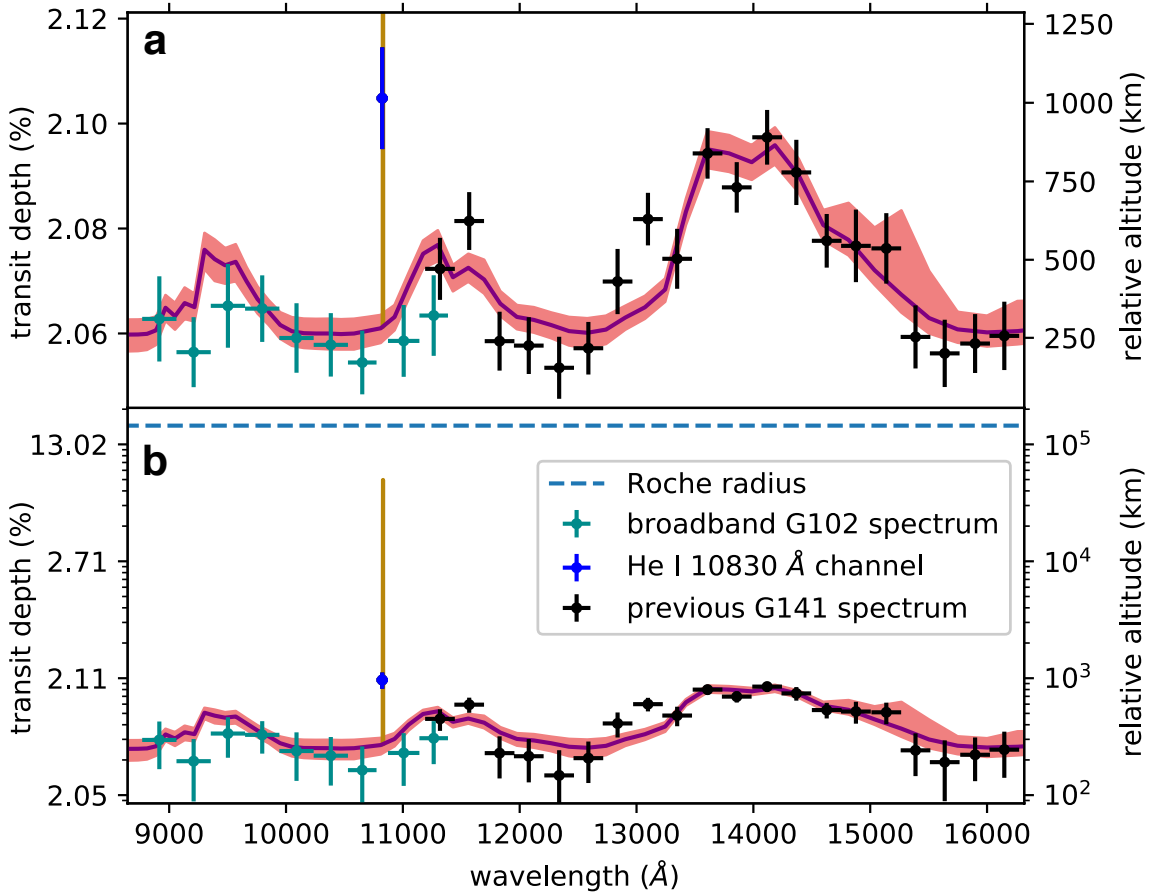

**Figure 1 | Combined near-infrared transmission spectrum for WASP-107b with helium absorption feature.** (a) Data plotted on a linear scale. Points with 1σ error bars are from a previous study[18] and this work, both corrected for stellar activity (see Methods). The solid purple line is the best fit lower atmosphere retrieval model from MCMC fits, and the shaded pink areas encompass 68%, 95% and 99.7% of the MCMC samples. The gold line is the best-fit helium 10,830 Å absorption profile from our 1-D escaping atmosphere model. (b) Same as (a), on a log scale. The dashed blue line shows the Roche radius.

The broadband transmission spectrum is consistent with a previous transmission spectrum for WASP-107b obtained using the WFC3 G141 grism, which covers the 11,000-16,000 Å wavelength range[18]. The latter exhibits a muted water absorption band centred at 14,000 Å, with an otherwise flat spectrum implying an opaque cloud deck. After applying a correction for stellar activity variations between the G102 and G141 observation epochs (see Methods), the G102 spectrum aligns with the cloud deck level inferred from the G141 spectrum (Fig. 1). The helium triplet has an expected width of approximately 3 Å, whereas the resolution of the G102 grism is 67 Å (~3 pixels) at 10,400 Å[19]. Therefore, to make a finely-sampled transmission spectrum, we shifted each of the 20 narrowband channels by 1 pixel with respect to the adjacent channel along the wavelength axis. The narrowband transmission spectrum peaked when the binning was most closely centred at 10,833 Å (Fig. 2), as expected if absorption by helium in the planetary atmosphere was responsible for the signal. To estimate the amplitude of the absorption feature, we focussed on 5 non-overlapping channels centred on 10,833 Å. All but one of the channels were consistent with a baseline transit depth level of 2.056 ± 0.005 %. The single exception is the channel centred on the 10,833 Å helium triplet, for which the transit is visibly deeper than for the surrounding channels (Fig. 3), and we obtained $(R_p/R_s)^2$=2.105 ± 0.010 %. We ruled out various alternative explanations for the signal, including other absorbing species, helium in the Earth's atmosphere, and the occultation of inhomogeneities in the stellar chromosphere and photosphere (see Methods).

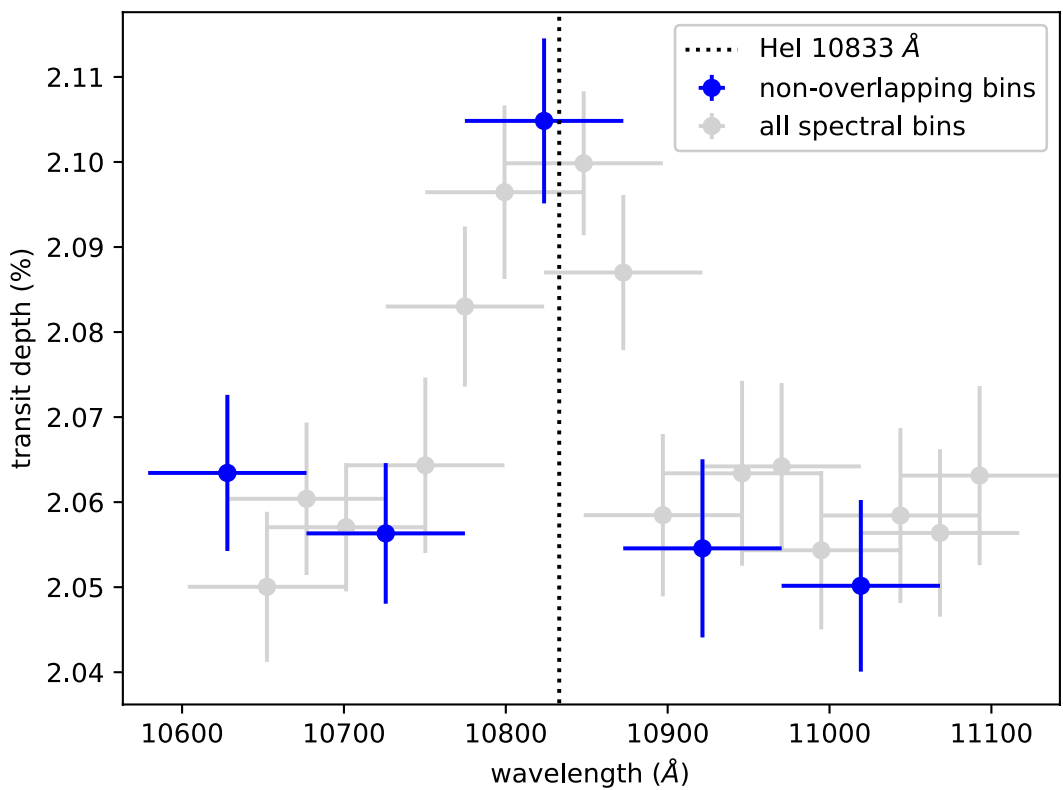

**Figure 3 | Narrow-band transmission spectrum of WASP-107b, centred on 10,833 Å.** Each spectroscopic channel has been shifted along one pixel from the last. Non-overlapping bins are highlighted in blue. Error bars are 1σ. The peak of the spectrum coincides with the $2^3S$ helium absorption line at 10,833 Å.

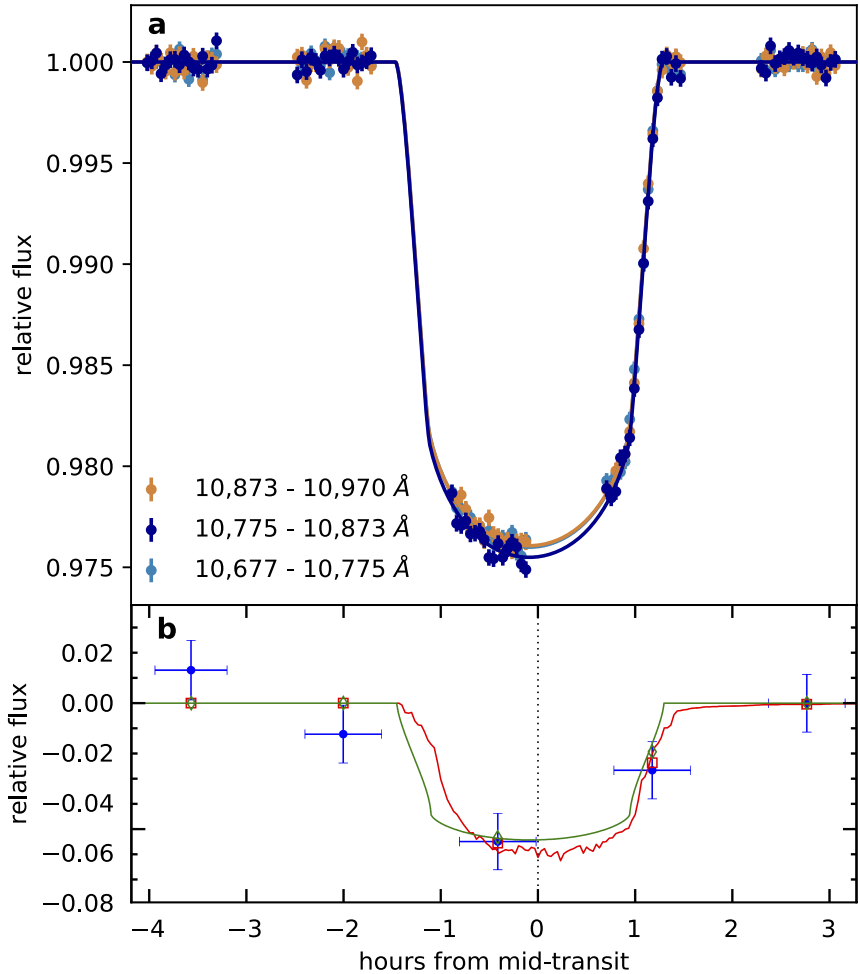

**Figure 2 | Transit light curves for three 98 Å -wide spectroscopic channels.** (a) Dark blue points are from the channel centred on the He I 10,833 Å line, gold and light blue points are from the two adjacent channels. All have 1σ error bars. The transit depth of the blue light curve is visibly deeper. (b) Binned difference between the 10,775 – 10,873 Å channel light curve, and the average of the two adjacent channels (blue points, 1σ errors), highlighting the excess absorption. It is well explained by both our 1D (green line) and 3D (red line) escaping atmosphere models.

The metastable helium probed by 10,833 Å absorption forms high up, at nanobar - μbar pressures in planetary atmospheres, where stellar XUV radiation is absorbed[12]. On the other hand, absorption of the neighbouring continuum occurs deeper in planetary atmospheres, at mbar - bar pressures. Therefore, to interpret the broadband (continuum) and narrowband (~10,833 Å) transmission spectra, we used separate lower- and upper- atmosphere models. For the combined G102 and G141 broadband spectrum (with the 10,775 - 10,873 Å range removed), we performed an atmospheric retrieval analysis using our one-dimensional radiative transfer code, ATMO[20,21] (see Methods and Extended Data Table 3). We found the broadband data were well explained by a grey absorbing cloud deck across the full 8,780-

11,370 Å wavelength range, in addition to H$_2$O absorption. We obtained a volume mixing ratio for H$_2$O of $5\times10^{-3} - 4\times10^{-2}$, consistent with previous estimations[18].

We investigated the narrowband transmission spectrum using two numerical models for the upper atmosphere of WASP-107b (see Methods). Our first, 1-D model[22] solves for the level populations of a H/He Parker wind, and suggests that WASP-107b is losing its atmosphere at a rate of $10^{10} - 3\times10^{11}$ g s$^{-1}$, corresponding to ~0.1 - 4% of its total mass every billion years. Our second, 3-D model[8,23] suggests an escape rate for metastable helium of $10^6$-$10^7$ g/s (for comparison, the 1-D model gives an escape rate of ~$10^5$ g s$^{-1}$ for $2^3$S helium). It also suggests that stellar radiation pressure blows away the escaping helium atoms so swiftly as to form a tail nearly aligned with the star-planet axis, which explains the lack of post-transit occultation detected in our data (Figure 2). The radiation pressure should also blue-shift the absorption signature over hundreds of km s$^{-1}$, which may be observable at higher spectral resolution (Fig. 4).

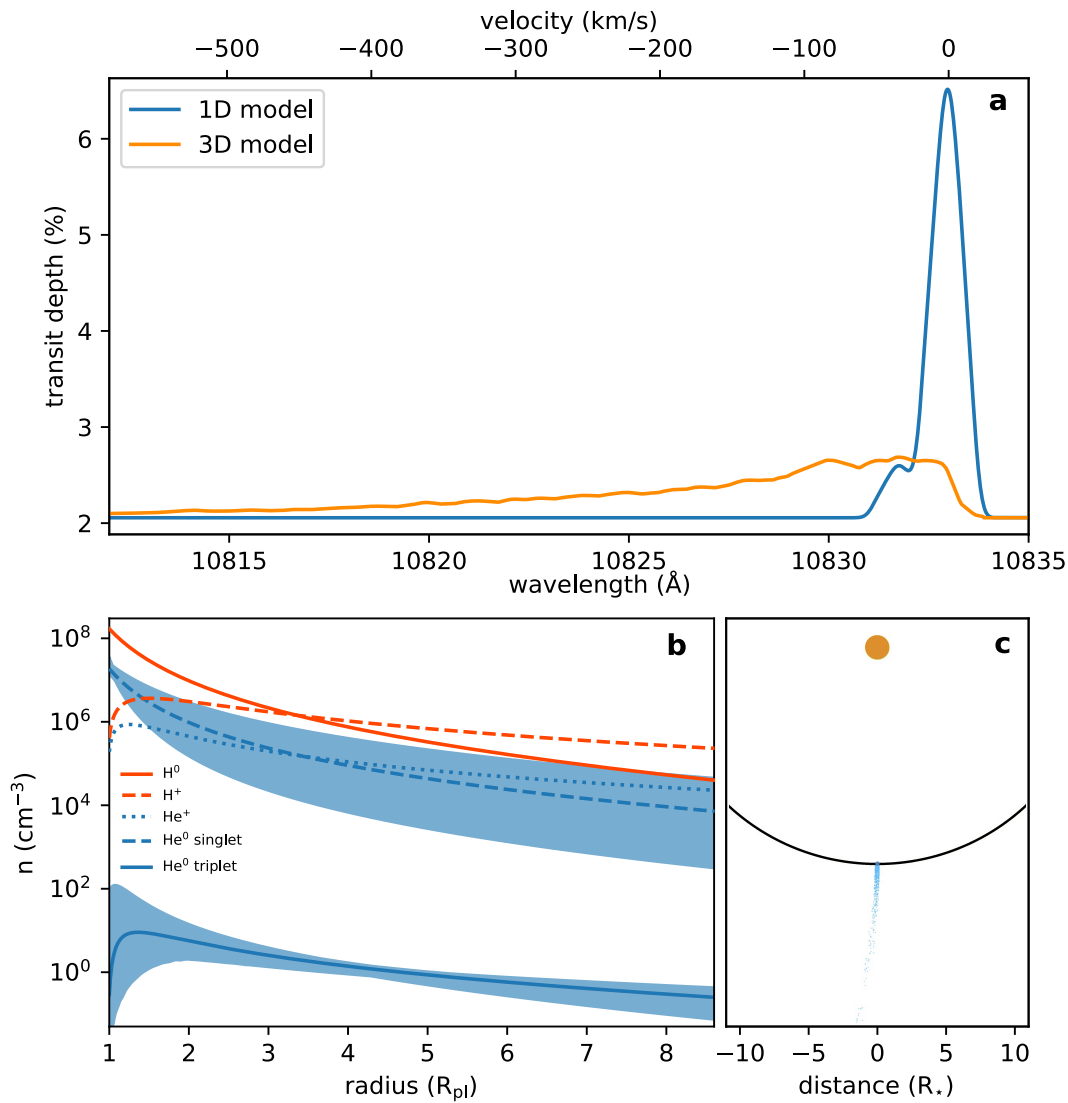

**Figure 4 | Results from two models of WASP-107b's upper atmosphere.** (a) Best-fit absorption profiles of the helium 10,833Å triplet line from the 1-D (blue), and 3-D (orange) models. Both reproduce the measured excess absorption of 0.049±0.011% in a 98 Å bin. Higher-resolution observations will resolve the profile shape, and further constrain the

velocity of the planetary wind. (b) Radial number density profiles of different atmospheric species from the 1-D model, shaded blue regions are 1σ errors. (c) Top-down view of the planetary system from the 3-D model, showing a comet-like tail of $2^3$S helium shaped by stellar radiation pressure.

Atmospheric mass-loss can substantially alter the bulk composition of a planet. For example, there is evidence that atmospheric escape is responsible for the observed dearth of highly-irradiated super-Earth and sub-Neptune exoplanets with sizes between 1.6 and 2 Earth radii[24-28]. In order to calibrate theories of planet formation, and assess whether these planets have substantial H/He envelopes, it is necessary to understand how atmospheric mass-loss affects the subsequent evolution of bodies that start with significant atmospheres. Empirical constraints such as the one presented here for WASP-107b are therefore crucial for retracing evolutionary pathways and interpreting the present day population of planets[29].

To date, extended atmospheres have been detected on three exoplanets by targeting the Lyman-alpha line in the UV[4,7,8], and on one exoplanet using the optical H-alpha line[11]. Our observations of WASP-107b in the 10,833Å line provide not only the first detection of helium on an exoplanet, but also the first detection of an extended exoplanet atmosphere at infrared wavelengths. This result demonstrates a new method to study extended atmospheres which is complementary to the two hydrogen lines.

We note that observations targeting the 10,833 Å helium triplet are possible from the ground with existing high-resolution infrared spectrographs. In the near future, high signal-to-noise observations will also be possible with the James Webb Space Telescope at a spectral resolution of $\Delta\lambda \sim 4$ Å ($\sim 110$ kms$^{-1}$).

**Online Content** Methods, along with any additional Extended Data display items and Source Data are available in the online version of the paper; references unique to those section appear only in the online paper.

**Received ;**

**Acknowledgements** We thank S. Seager, A. Dupree, V. Andretta, M. Giampapa and B. Drummond for discussions. This work is based on observations with the NASA/ESA *HST*, obtained at the Space Telescope Science Institute (STScI) operated by AURA, Inc. J.J.S. is supported by an STFC studentship. The research leading to these results has received funding from the European Research Council under the European Union's

Seventh Framework Programme (FP7/2007-2013) / ERC grant agreement no. 336792. Support for this work was provided by NASA through grants under the HST-GO-14916 programme from the STScI. G.W.H. and M.H.W. acknowledge support from Tennessee State University and the State of Tennessee through its Centers of Excellence program. The MEarth Team gratefully acknowledges funding from the David and Lucille Packard Fellowship for Science and Engineering, the National Science Foundation, and the John Templeton Foundation. The opinions expressed in this publication are those of the authors and do not necessarily reflect the views of the John Templeton Foundation. This work has been carried out in the frame of the National Centre for Competence in Research PlanetS supported by the Swiss National Science Foundation (SNSF). VB, DE, AW and SU acknowledge the financial support of the SNSF. DE and VB acknowledge funding from the European Research Council (ERC) under the European Union's Horizon 2020 research and innovation programme (project FOUR ACES; grant agreement No 724427)

**Author Contributions** J.J.S. led the HST telescope time proposal, designed the observations, and led the data analysis with contributions from T.M.E., H.R.W., L.K and Y. Z.. J.J.S. identified the planetary helium, and wrote the manuscript with contributions from T.M.E., V.B., A.O., J.I., B.V.R and G.W.H. A.O. and V.B. performed detailed modelling of the exosphere, with contributions from D.E. D.K.S. provided scientific guidance and performed the retrieval analysis. J.I., G.H., M. H. and D.C. provided ground-based photometry to correct for stellar activity. All authors discussed the results and commented on the paper.

**Author Information** Reprints and permissions information is available at www.nature.com/reprints. Readers are welcome to comment on the online version of the paper. Correspondence and requests for materials should be addressed to J. J. S. (jspake@astro.ex.ac.uk)
**Competing interests**
The authors declare no competing interests.

## Methods

### Observations & data reduction

We observed one transit of WASP-107b with WFC3 in spectroscopic mode, using the G102 grism (GO-14916, P.I. Spake). This covers the approximate wavelength range of 8,780 – 11,370 Å. We used forward spatial scanning to spread the spectra over ~60 pixels in the cross-dispersion direction with the SPARS10, NSAMP=15 setup, giving exposure times of ~103 seconds. This allowed 17 exposures per HST orbit. The observations lasted for five HST orbits, with two orbits pre-transit, one during the transit, and one post-transit, allowing us to precisely constrain the out-of-transit baseline.

The raw frames were first reduced with the automatic CalWF3 pipeline. The 1-D spectra were then extracted following standard methods[30]: building up flux counts by summing the difference between successive non-destructive reads. We removed the background from each read difference by subtracting the median of a box of pixels uncontaminated by the spectrum. We found the flux-weighted centre of each scan and set to zero all pixels more than 75 rows away from the centre in the cross-dispersion axis, which removes many cosmic rays. The remaining cosmic rays were flagged by finding 4$\sigma$ outliers relative to the median along the dispersion direction. We replaced each flagged pixel with the median along the dispersion direction, re-scaled to the count rate of the cross-dispersion column. Since the scans are visibly tilted from the dispersion axis, we used the IRAF package Apall to fit the trace of the 2-D scans and extract 1-D spectra. We found the wavelength solutions by cross-correlating the extracted spectra with an ATLAS model stellar spectrum[17] which most closely matches WASP-107 ($T_{eff}$ = 4,500 K, log $g$ = 4.5 cgs) modulated by the G102 grism throughput. Following standard methods[18] we interpolated each spectrum onto the wavelength range of the first to account for shifts in the dispersion axis over time.

### White light curve analysis

We extracted the white light curve by summing the total counts of each 1-D spectrum. In order to constrain the mid-time of the transit, we fit the resulting time-series with the BATMAN transit model[14], multiplied by a linear baseline trend and a physically-motivated systematics model. For the latter, we employed the RECTE model[15], which accounts for two populations of charge traps in individual pixels of the detector and successfully replicates the ramp-like features that dominate the systematics. The RECTE model allows us to keep the first orbit of observations in our fit. The free parameters of our final model were: the planet-to-star radius ratio, $R_p/R_s$; mid-transit time, $T_0$; the gradient and y-intercept of the linear background trend, $c_1$ and $c_0$ respectively; four parameters for the charge trapping model - the initial number of populated slow and fast traps $s_{pop}$ and $f_{pop}$, and the changes in the two populations between each orbit, $\delta s$ and $\delta f$; and an uncertainty rescaling factor, $\beta$ for the expected photon noise.

We fixed $a/R_s$, $i$, $e$, and the period using estimates from *Kepler* light curves[16]. To model the stellar limb darkening we fitted a four-parameter non-linear limb darkening law[31] to the ATLAS stellar model described above.

Because the shape of the ramp-like systematics depends on the count level of the illuminated pixels, the RECTE model requires the `intrinsic' count rate of a pixel (i.e. the actual flux received from the star) in order to model the charge trapping. To create a template of the intrinsic count rate, we median-combined four raw images from the end of the second orbit. Here the charge traps appear completely filled, and the ramp shape has tapered to a flat line. It is possible to model each illuminated pixel, however, for a large scan this is computationally expensive. Additionally, the ramp profile is washed out by systematics that are introduced by telescope jittering and pointing drift. Instead we divided the scan into columns of width 10 pixels along the dispersion axis and fed the median count profiles into the model.

We used the Markov chain Monte Carlo (MCMC) package *emcee*[32] to marginalise over the parameter space of the model likelihood distribution. We used 80 walkers and ran chains for 8000 steps, discarding the first 800 as burn-in before combining the walker chains into a single chain. The best-fit model and residuals are shown in Extended Figure 1, with the parameter values and 1σ uncertainties reported in Extended Data Table 1. Although WASP-107b orbits an active star we see no evidence of star spot crossings. For context, only five spot-crossing events are reported in 10 *Kepler* transits[16,33].

**Broadband spectroscopic light curve fit**

We binned each spectrum into nine spectroscopic channels across the 8,780-11,370 Å wavelength range, each spanning 10-12 pixels on the detector. The resulting lightcurves are shown in Extended Data Figure 2. Since the throughput of the G102 grism is wavelength-dependent, the shape of the charge-trapping ramp in each spectroscopic light curve is different. Therefore, for each channel we simultaneously fit for a transit model multiplied by a linear baseline trend and a charge-trap model. To make a template of the intrinsic counts, we took the median cross-dispersion-direction profile of each channel in the same four raw images as used in the white light curve fit. We fixed $T_0$ to the value found from the white light curve fit. Similarly to the white light curve fit, we fixed the orbital parameters to those derived from *Kepler* light curves[16], and wavelength-dependent limb darkening coefficients from the ATLAS model. Therefore, for each channel the fitted parameters were $R_P/R_s$, $c_1$, $c_0$, $s_{pop}$, $f_{pop}$, $\delta s$, $\delta f$, and $\beta$. We ran MCMC fits for each light curve with *emcee*, with 80 walkers, 80,000 steps and a burn-in of 800.

As a test, we also ran additional fits for the spectroscopic light curves with the stellar limb darkening coefficients as free parameters. This produced results that were consistent to within 1σ with those obtained from the analysis in which the limb darkening coefficients were held fixed.

We show the resulting spectroscopic light curves divided by their best-fit systematics models in Extended Data Figure 2, along with their residuals. Extended Data Table 2 reports our median values for the transit depth, $(R_P/R_s)^2$, with $1\sigma$ uncertainties calculated from the MCMC chains. We also list the root mean square (RMS) of the residuals for each channel, which range between 1.038-1.198 times the photon noise.

**Narrowband spectroscopic light curve fit around 10,830 angstroms**

To target the 10,833 Å helium triplet, we binned the spectra from 10,590 to 11,150 Å into twenty narrowband channels. Each channel spanned 4 pixels on the detector, which is a compromise between the low instrument resolution, signal-to-noise, and the narrowness of the targeted feature. The wavelength coverage of each channel was shifted relative to the adjacent channel by one pixel, so the channels overlap.

We note that since the formal resolution of the G102 grism is $\lambda/\Delta\lambda \sim 155$ at 10,400 Å[19] (which corresponds to $\Delta\lambda \sim 67$ Å, or 2.7 pixel widths), the smallest bins theoretically possible are 3 pixels wide. A resolution of 3 pixels could be achieved if the 10,833 Å feature lay in the centre of a pixel, but in our data it lies significantly blue-ward of the centre of its pixel. This means there is some 10,833 Å flux in the pixel located two pixels blueward of the 10,833 Å line. Indeed, when we tested the 3-pixel case we found that the amplitude of the 10,833 Å feature increased by 0.011% from the 4-pixel-bin fit, which is similar to the expected increase of 0.016% if all the 10,833 Å flux fell within a central 3-pixel bin. With 3-pixel bins the feature also appeared to have a slight blue 'wing', which is unlikely to be astrophysical, as such wings would be expected from binning the data to a resolution higher than that of the spectrograph. We therefore used conservative 4-pixel bins.

Extended Data Figure 3 shows the spectroscopic light curves divided by their best-fit systematics models, along with their residuals. Extended Data Table 2 shows our median values for the transit depth and their $1\sigma$ uncertainties, calculated from the MCMC chains. We also list the RMS of the residuals of each channel, which range from 0.976 to 1.22 relative to photon noise. The resulting transmission spectrum is shown in Figure 2.

Previous studies[34] have highlighted the importance of considering the effect of stellar limb darkening in stellar absorption lines on exoplanet transmission spectra. To investigate whether this could cause the strong feature at 10,833 Å, we re-ran the narrow-band spectroscopic light curve fits whilst fitting for a quadratic limb-darkening law. The resulting spectrum was consistent with our previous analysis within 1-σ.

Strong stellar lines that shift over the edges of pixels can introduce noise to measured transmission spectra[35]. We checked this effect by smoothing our extracted time series spectra with a Gaussian kernel of FWHM of 4 pixels, and re-running the narrowband spectroscopic light curve fits. Our measured 10,833 Å absorption feature remained consistent within 1 σ.

**MEarth observations**

Photometric monitoring observations were gathered using a single telescope of the MEarth-South[36,37] array (CS 2015) at Cerro Tololo Inter-American Observatory (CTIO), Chile. Data were obtained on 78 nights from 2017 March 22 (UT) to 2017 August 1 in groups of 4 × 15s exposures, with these exposure groups repeated at a cadence of approximately 30 minutes. A total of 3096 exposures were gathered over this period. The bandpass of these observations is in the red optical with the blue cutoff defined by RG715 glass at approximately 7,150 Å and the red cutoff defined by the decline of the CCD quantum efficiency at approximately 10,000 Å. For our data reduction, we used our previously published methodology[38], modified for the specifics of the MEarth data[39].

The CCD camera shutter failed on 2017 May 9, which required removal for servicing. This procedure introduces flat-fielding errors not corrected to sufficient precision by standard calibrations, so instead we allow for this explicitly in the analysis by solving for a change in the magnitude zero-points on both sides of the meridian at this date, following standard methods[40]. The result of this analysis is a "least-squares periodogram" (shown in Extended Data Figure 4), obtained by simultaneously fitting a periodic modulation, while accounting for the four magnitude zero-points and two additional linear terms describing sources of systematic errors in the photometry (FWHM of the stellar images and the "common mode" as a proxy for the effect of variable precipitable water vapor on the photometry). This procedure would be mathematically equivalent to a Lomb-Scargle periodogram in the absence of these six extra terms. The highest peak in the periodogram and its full width at half-maximum corresponds to a periodicity of 19.7±0.9 days. This is consistent with estimates from Kelper light curves of 17.5±1.4 days[33]. We find an amplitude of ~0.00150 in magnitude.

**AIT Photometry**

We acquired nightly photometric observations of WASP-107 with the Tennessee State University Celestron 14-inch (C14) automated imaging telescope (AIT) located at Fairborn Observatory in southern Arizona[41,42]. The observations were made in the Cousins R passband with an SBIG STL-1001E CCD camera. Differential magnitudes of WASP-107 were computed with respect to eight of the most constant comparison stars in the CCD field. Details of our data acquisition, reduction, and analysis can be found in a previous work[43], which describes a similar analysis of the planetary-host star WASP-31.

A total of 120 nightly observations (excluding a few observations in transit) were collected between 2017 Feb. 23 and June 28. The nightly differential magnitudes are plotted in panel (a) of Extended Data Figure 5. Panels (b) and (c) show the frequency spectrum of the observations and the phase curve computed with the best frequency. Our frequency analysis is based on least-squares sine fits with trial frequencies between 0.01 and 0.5 c/d, corresponding to periods between 2 and 100 days. The goodness of fit at each frequency is measured as the reduction factor in the variance of the original data. Low-amplitude brightness variability is seen at a period of 8.675±0.043 days with a peak-to-peak amplitude of only 0.005 mag. Our period is almost exactly half the 17.5-day rotation period found in *Kepler* light curves[33] and demonstrates that WASP-107 has spots or spot groups on opposite hemispheres of the star during the epoch of our observations. The WASP-107b discovery team[6] also found periods of around 17 and 8.3 days in their 2009 and 2010 photometry.

**Stellar variability correction**

To correct for stellar variability between the G141 and G102 epochs, we follow a similar method to previous studies[44,45], and estimate the flux from the non-spotted stellar surface as $F_s = max(F) + k\sigma$, where $F$ is the photometric light curve, $k$ is a fitted value and $\sigma$ is the scatter of the light curve. A previous study[44] found that $k = 1$ is a good value to use for active stars, so we adopt this value. We use the best-fit period, amplitude and ephemeris from the MEarth photometry to estimate the expected flux dimming correction at the mid-transit times for both data sets. We used the wavelength-dependent spot correction factor developed in a previous work[46] to correct for unoccluded spots, and we set the spot temperature to be 3200K. After the correction, the two spectra align well and appear to share a flat baseline. The one overlapping spectral channel between G102/G141 is consistent within 1$\sigma$.

**ATMO retrieval**

For the combined G102 and G141 broadband spectrum corrected for photoshperic variability, we performed an atmospheric retrieval analysis using our one-dimensional radiative transfer code, ATMO[20,21,47,48,49]. We assumed an isothermal temperature-pressure profile, and used MCMC to fit for the following parameters: atmospheric temperature; planetary radius at a pressure of 1 mbar; grey cloud opacity; and the abundances of $H_2O$, $CO_2$, CO, $CH_4$, $NH_3$, $H_2S$, HCN and $C_2H_2$. We assumed solar abundances under chemical equilibrium for other gas species. Note that for this analysis we excluded wavelengths coinciding with the narrowband channel centred on the 10,833 Å helium triplet. Our best-fit model is shown in Figure 1, with a a $\chi^2$ of 31.4 for 18 degrees of freedom.

**Assessing detector defects and random noise**

We checked that the residuals for the pixel columns in each frame do not reveal any obvious anomalies over the narrow 10,833 Å helium triplet, which suggests that it is not caused by a detector defects or uncorrected cosmic rays. In addition, the transit depths remained consistent within 0.5σ when we removed 1/3 of the points in the light curves, in several random sub-sets, and re-fit them with the same procedures as described above.

**Absorption from other species**

The strong absorption line of metastable $2^3$S helium at 10,833 Å aligns extremely well with the peak of the feature. In the 20 Å region surrounding this peak (10,820 to 10,840 Å), helium is the only species that contains absorption solely within this wavelength range but nowhere else within the G102 bandpass (8,060 to 11,170 Å). There is, for example, a strong silicon absorption line at 10,830 Å [50], and a water line at 10,835 Å (vacuum wavelengths) [50], but if either species were the cause of the absorption seen in our transmission spectrum, there would be other similarly strong silicon lines measured at 10,588, 10,606 and 10,872 Å, and a water line at 10,929 Å, where we see no excess absorption. The other atoms with strong absorption lines near 10,833 Angstrom are Np, Cs, Fe, Th, S, Cr, V, Yb, and Cu – all of which can be ruled out as they are either radioactive with short half-lives, or have other strong transitions within the the 8,060 to 11,170 Å wavelength range that we do not observe. We have also found there to be no species in the ExoMol[51] or HITRAN/HITEMP[52,53] databases with sufficiently sharp features aligned at 10,833 Å. Specifically, we searched the following species: $CH_4$, $CO_2$, HCN, NH, CH, OH, PO, NO, VO, TiO, CN, $C_2$, $PH_3$, $NH_3$, SiO, CaO, $H_3+$, CO, $H_2CO$, $C_2H_2$, BeH, LiH, HCl, AlO, $SO_2$, $H_2S$, PN, KCl, NaCl, CS, CP, PS, MgH, NaH, CrH, CaH, FeH, and ScH. We therefore conclude that absorption by metastable helium at 10,833 Å is the most plausible explanation for the signal detected in the narrowband transmission spectrum.

**Assessing the Earth's exosphere**

Where the Earth's exosphere is illuminated by XUV radiation from the sun, there is metastable helium. At an altitude of ~500km, HST passes right through the Earth's exosphere, and when not in the Earth's shadow, will pass through regions containing metastable helium. The change in abundance of the metastable state throughout orbit has been shown to impart time-varying background signal in the 10,833 Å line on the timescale of one ~95 minute spaceraft orbit[54]. There is no telluric metastable helium in Earth's shadow, and as expected, there is no significant excess absorption at 10,833 Å while HST is in Earth shadow[54]. It does, however, affect HST measurements at dawn and dusk - i.e. when the spacecraft passes through the solar-illuminated upper atmosphere. The magnitude of the effect is correlated with the solar activity cycle – i.e. more activity, more UV, more metastable helium. The effect of spatially-diffuse telluric helium emission on WFC3 slitless spectroscopy is to impart an increased sky background signal across the detector. At the time of the observations, we were approaching solar minimum, and the 10.7cm radiation (which is a proxy for solar activity) was only 70 solar flux units, sfu (Solar Monitoring Program, Natural Resources Canada). According to the WFC3 instrument report[54] observations only appear significantly affected when the 10.7cm flux is greater than ~100 sfu. Nonetheless, to test whether metastable helium at dawn and dusk in the Earth's atmosphere could cause an anomalous absorption feature in our transmission spectrum, we removed the first and last 4 exposures of each orbit – which encompasses the initial and final 10 minutes - when HST passed through the illuminated dusk and dawn exosphere, and re-fit the light curves. The results were consistent with previous analysis at less than 1 σ, which indicates that emission from telluric helium is not the cause of the narrowband absorption feature in our data. We note that previous transit spectroscopic studies using G102[55,56] do not show excess absorption at 10,833 Å.

**Assessing the stellar chromosphere**

We also considered the possibility that the absorption feature we measure at 10,833 Å could be a result of stellar activity, since the metastable $2^3S$ state of helium is formed in the inhomogeneous upper chromospheres and coronae of stars via photo-ionisation, recombination, and collisional excitation. The planet passing over quiet regions with less 10,833 Å helium absorption could in theory increase the relative transit depth at this wavelength and thus mimic an exoplanet atmospheric feature.

Theoretical models of chromospheres[57,58] predict the maximum equivalent width of the 10,833 Angstrom helium line in the spectra of F- to early K-type stars to be ~0.4 Å. Being a K6 star, WASP-107 lies just outside the valid range of spectral types for this model. However, in the following section we show that in order to match our observed transmission spectral feature, the nominal chromospheric absorption at 10,833 Å of the WASP-107 host star would need to be five times stronger than any isolated (i.e. non-multiple), main-sequence dwarf star measured to date.

After searching the literature for all 10,833 Å helium triplet equivalent width measurements of isolated dwarf stars, we found over 300 measurements of over 100 distinct stars, including 23 measurements of 11 different stars of similar spectral type to WASP-107 (K5-K7). We found no measurements greater than 0.409 Å[59-64]. We took an additional measurement of the K6 star GJ380 with NIRSpec on Keck, which was found to have an equivalent width of 0.311 Å (A. Dupree, private communication).

Furthermore, it has been shown[55,63] that the equivalent width of the 10,833 Å line is related to that of another neutral helium absorption line, at 5,876 Å. The 5,876 Å line is produced by the transition from the $2^3D$ to the $2^3P$ state. As such, the 5,876 Å line forms in the same regions of the stellar chromosphere as the 10,833 Å triplet (which corresponds to the $2^3S$ to $2^3D$ transition). Extended Data Figure 5 shows the equivalent width measurements of the 10,833 and 5,876 Å lines in a survey of 31 FGK stars[63]. A strong correlation is apparent. To investigate the 5,876 Å helium line of WASP-107, we co-added high-resolution spectra obtained with the HARPS spectrograph (ESO program 093.C-0474(A)). These spectra cover a wavelength range of 3,800 to 6,900 Å (Extended Data Figure 5). We fit for the equivalent width of the 5,876 Å helium line in the co-added spectrum, with the result indicated on Extended Data Figure 6 as a yellow shaded region. We find the equivalent width of this feature is similar to that measured for other single dwarf stars, with no evidence of unusual activity. Given the well-established correlation between the equivalent widths of the 5,876 and 10,833 Å helium lines noted above, this is further evidence against the WASP-107 host star having an abnormally deep 10,833 Å line. In addition, we measured the S-index for WASP-107 from the HARPS spectra, and found a night-averaged value of $S_{HK}=1.26\pm0.03$ (A.W., private communication), which is a moderate value for a K6 star[64].

We therefore adopt the maximum equivalent width of 0.4 Å to estimate an upper limit for the amplitude of a feature that could be caused by un-occulted 10,833 Å helium absorption of stellar origin in our 98- Å -wide spectroscopic channel. We consider the limiting case in which WASP-107b occults only quiet regions of the star, where we assume there is no 10,833 Å absorption. This is the scenario in which the maximum amount of stellar continuum flux at 10,833 Å would be blocked out by the planet, which we treat as a fully opaque disk. We estimate the increased transit depth to be

$$D_{activity} = \frac{A_{pl}}{1 - \frac{W_{He}}{W_{bin}}} = 2.064 \pm 0.005\%$$

where $A_{pl}$=2.056±0.005% is the fraction of the stellar area occulted by the planet; $W_{He}$= 0.4 Å, is the maximum equivalent width of the stellar absorption feature; and $W_{bin}$ is the width of the spectral bin (i.e. 98 Å). This gives an upper limit of the feature caused by stellar activity, $\delta D_{activity} = D_{activity} - A_{pl} = 0.008±0.005\%$, which is less than one fifth of the measured size of the feature (0.049±0.011%). We therefore conclude that the observed absorption feature cannot be caused by stellar chromospheric spatial inhomogeneity alone.

**Resolution-Linked Bias**

If an absorption line overlaps in both a stellar and planetary atmosphere spectrum, and the line is unresolved in the measured transmission spectrum, then the planetary absorption can be underestimated. The effect is called Resolution Linked Bias (RLB)[65]. For the 10,833 Å line in the WASP-107 system this dilution effect will compete with the possible over-estimation of the signal from unoccculted chromospherically active regions (as described in the 'Assessing the stellar chromosphere' section). The magnitudes of both effects will depend on whether the planet transits in front of active or quiet regions of the star. The RLB effect would be largest if the planet transited only chromospherically active regions (which have the highest 10,833 Å absorption). We estimated the magnitude of the RLB effect in this limiting case following the method described in a previous work[65], and assuming an equivalent width of 0.4 Å for the 10,833 Å stellar line. For a measured absorption excess of 0.049±0.011% in a 98 Å bin centred on the 10,833 Å line, we could be underestimating the planetary absorption by up to 0.009% (i.e. about one fifth of the measured signal). However, without knowledge of which part of the chromosphere the planet transits; the stellar line profile; and the velocity structure of the planetary helium signature, we cannot accurately estimate the magnitudes of the competing effects.

**Stellar flares**

The He 10,833 Å line appears in emission in solar- (and presumably stellar-) flares[66], so active stars like WASP-107 could show short-term variability in the line, which may be difficult to disentangle from a transiting planetary signal[34]. Flares are unlikely to wholly mimic the signal we detect, since the planet would need to pass in front of flaring regions of the star throughout the duration of the transit. Instead, unoccculted flares could dilute He 10,833 Å atmospheric absorption. Visual inspection of the raw light curve of the spectroscopic channel centred on 10,833 Å shows no evidence of flare events. Additionally, the pre- and post- transit flux levels agree with each other, which would not be the case if there was significant 10,833 Å emission from the tail of a flare. As a precaution, we re-produced the narrowband transmission spectrum around the 10,833 Å line using different combinations of the out-of transit baseline: firstly with only orbits 2 and 4, then with orbits 1 and 3, and then orbits 2 and 5. All three cases gave a 10,833 Å absorption feature consistent to within 1σ of our full fit.

**Photospheric spots and faculae**

To quantify the effect of a heterogeneous photosphere on the transmission spectrum around 10,833 Å, we used a variability modelling method[67,68] which uses an ensemble of model stellar photospheres with randomly located active regions to provide estimates of the fraction of the stellar surface covered by photospheric spots and faculae for a given rotational variability amplitude. While variability monitoring traces only the non-axisymmetric component of the stellar heterogeneity and thus provides a lower limit on active region covering fractions[68], this numerical approach provides a more complete understanding of the range of covering fractions that may correspond to an observed variability level. The model describes the integrated full-disk spectrum by the combination of three components: the immaculate photosphere, spots, and faculae. We used three spectra interpolated from the PHOENIX model grid[69] with log g = 4.5 and [M/H] = +0.02 and different temperatures to represent the three components. Following previous works[68], we set the photosphere temperature, $T_{phot}$, to the effective temperature of the star ($T_{eff}$=4430 K[6]) and adopt scaling relations for the spot temperature $T_{spot}$[70,71] and faculae temperature $T_{fac}$[72].
Thus, the temperatures of the three components are $T_{phot} = T_{eff} = 4,430$ K, $T_{spot} = 0.73 \times T_{phot} = 3,230$ K, and $T_{fac} = T_{phot} + 100$ K $= 4,530$ K. WASP-107b's discovery paper[6] reports a 17-day periodic modulation with a 0.4% semi-amplitude (0.8% full-amplitude) for WASP-107. Assuming a typical spot radius of $r_{spot} = 2°$, we find the reported rotational variability could be caused by a spot filling fraction of $f_{spot} = 4^{+9}_{-2}\%$ (1σ confidence interval) if the variability is due to spots alone. In the more realistic case in which spots and faculae are both contributing to the variability, we find $f_{spot} = 8^{+6}_{-3}\%$ and $f_{faculae} = 53^{+15}_{-12}\%$. The covering fractions we report are means over the entire model photosphere. They do not take into account relative over- or under-abundances of magnetic features on the Earth-facing hemisphere during a transit. Therefore, in the worst case scenario, they could underestimate the hemispheric covering fractions by a factor of 2. However, the 1-σ confidence intervals, which are derived from 100 model realizations with randomly selected active region locations, are deliberately conservative to account for this. Extended Data Figure 6 shows how unoccluded photospheric stellar heterogeneities could affect the transmission spectrum, assuming the planet transits a chord of immaculate photosphere. The stellar contamination factor, ε, on the y-axis is multiplied by the true $(R_p/R_s)^2$ transit depth to produce the observed transmission spectrum, i.e. $\varepsilon > 1$ means the observed transit depth is deeper than expected from the planetary atmosphere model. The spots+faculae model does not predict an increase in transit depth at 10,833 Å. No sharp features around 10,833 Å are apparent. Instead, the model predicts transit depths should be inflated by ~1% across the full wavelength range of G102 with perhaps some features apparent at ~8,500 Å and 8,900 Angstrom (for this reason we only use the 8,780-11,370 Å region in our full transmission spectrum, even though the G102 throughput extends down to 8,000 Å). The strong absorption feature we measure is therefore unlikely to be caused by photospheric inhomogeneity.

**1-D escaping atmosphere model**

Here we give a brief overview of the first model used to investigate the narrowband transmission spectrum at 10,833 Å, which is presented and described in more detail in a previous work[22]. This 1D model is based on the assumption that a thermosphere of a close-in exoplanet can be well represented by the density and velocity profile of an isothermal Parker wind driven by gas pressure[73]. We assume a composition of atomic hydrogen (90% by number) and helium (10%). We find the solution for the hydrogen ionization balance and the distribution of helium atoms in the ground, excited $2^3S$, and ionized states. The physical processes taken into account in the helium balance are photoionization from the ground and $2^3S$ states, recombination to the singlet and triplet states, collisional transitions between the triplet $2^3S$ state and states in the helium singlet ladder, which includes collisions with both free electrons and neutral hydrogen atoms, and the radiative decay from the $2^3S$ state to the ground state. The photoionization rates are calculated using the UV stellar flux of a K6 star HD 85512 taken from the MUSCLES survey[74] (version 2.1[75,76]), placed at the orbital distance of WASP 107b The equations used to compute the hydrogen/helium distributions, along with all the relevant reaction rate coefficients and cross sections, are described in a previous work[22]. We only changed the input parameters such as the mass and radius of the planet and its host star, as well as the input stellar spectrum, so that they match the properties of WASP 107b.

Based on the obtained density profile of helium in the $2^3S$ state, we calculate the optical depth and the in-transit absorption signal at 10,833 Å, assuming that a planet with a spherically symmetric thermosphere transits across the center of the stellar disk. For a planet of given mass and radius, the wind temperature and the total mass loss rate are free parameters in the model. Based on the results from the literature[77,78], we explore a temperature range between 5,000-13,000 K. In order to produce the absorption signal consistent with our measurement, the required mass loss rate is between $10^{10}$ and $3\times10^{11}$ g/s.

**3-D escaping atmosphere model**

Our second model has previously been used to interpret the escaping exosphere of the Neptune-mass exoplanet, GJ436b[9,23]. It considers neutral helium atoms that are released from the top of the thermosphere and subjected to planetary and stellar gravity, radiation pressure, and photoionization. We found that the data are well explained by $2^3$S helium atoms escaping at a rate of $10^6$-$10^7$ g/s. Stellar radiation pressure on the escaping helium atoms is stronger than the counter-balancing stellar gravity by a factor of approximately 10 and 50 for the weakest and strongest of the 10,833 Å triplet lines, respectively. Thus the gas blows away so swiftly as to form a tail nearly aligned with the star-planet axis.

**Code availability**

The custom code used to extract the HST spectra from the raw data frames is available upon request. The HST light curve fitting was performed using the open source BATMAN (https://github.com/lkreidberg/batman) and emcee codes (http://github.com/ dfm/emcee), and the proprietary RECTE code. The ATMO code used to compute the lower atmosphere models is currently proprietary, as are the 1-D and 3-D upper atmosphere codes.

**Data availability**

Raw HST data frames are publicly available online at the Mikulski Archive for Space Telescopes (MAST; https://archive.stsci.edu).

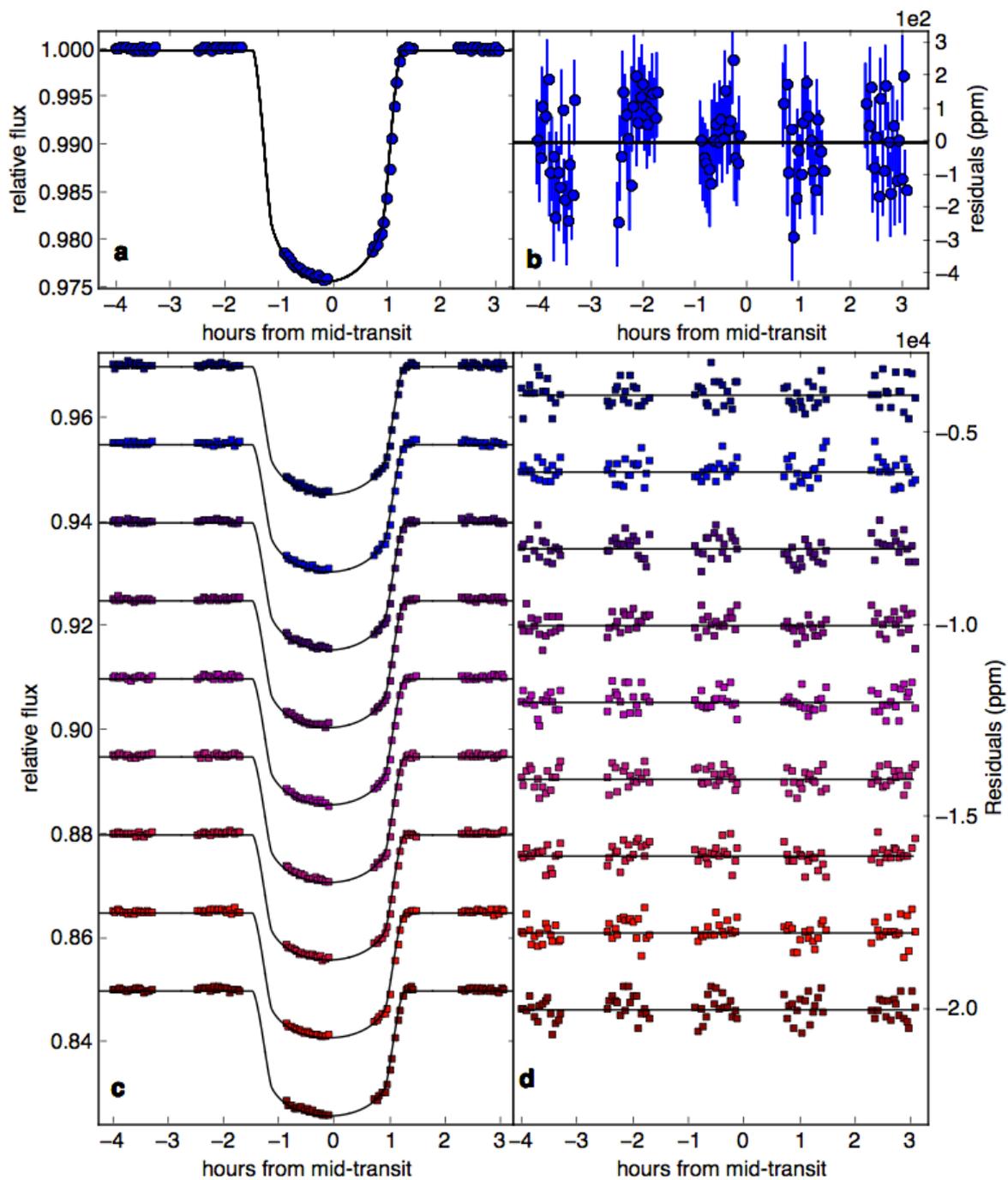

**Extended Data Figure 1 | G102 white light curve and broadband spectroscopic light curves covering the 0.88-1.14 micron wavelength range for WASP-107b.** (a) White light curve relative flux divided by systematics model, with best-fit transit light curve plotted in black. (b) White light residuals and 1σ errors, after removing the combined transit and systematics components of the best-fit model. (c) Points are spectroscopic light curves divided by systematics models, black curves are best-fit transit models, with vertical offsets applied for clarity. (d) Best-fit spectroscopic model residuals with vertical offsets applied for clarity.

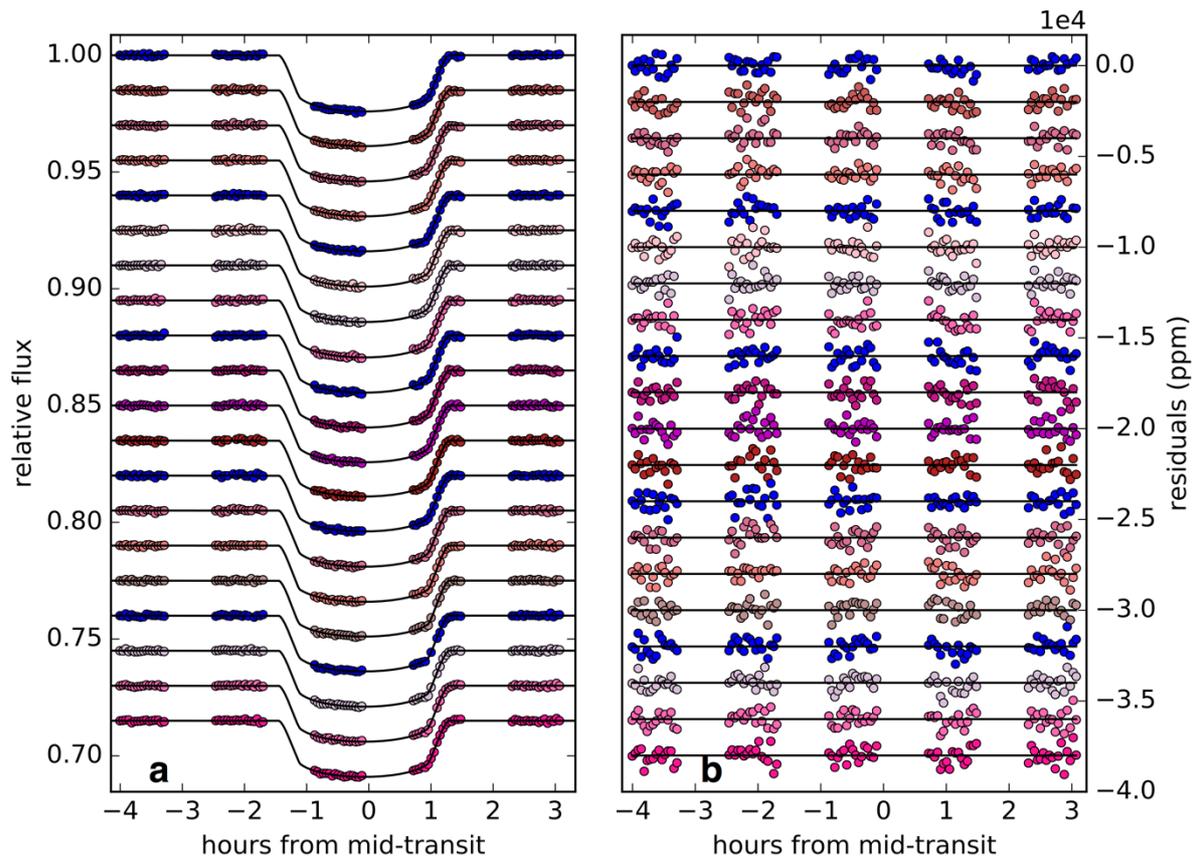

**Extended Data Figure 2 | Narrow-band (4-pixel-wide) spectroscopic light curves covering the 1.06-1.12 micron wavelength range.** (a) Points are light curves divided by systematics models, black curves are best-fit transit models. (b) Best-fit model residuals with vertical offsets applied for clarity. The 5 non-overlapping channels used to measure 10,833 Å absorption are highlighted in blue.

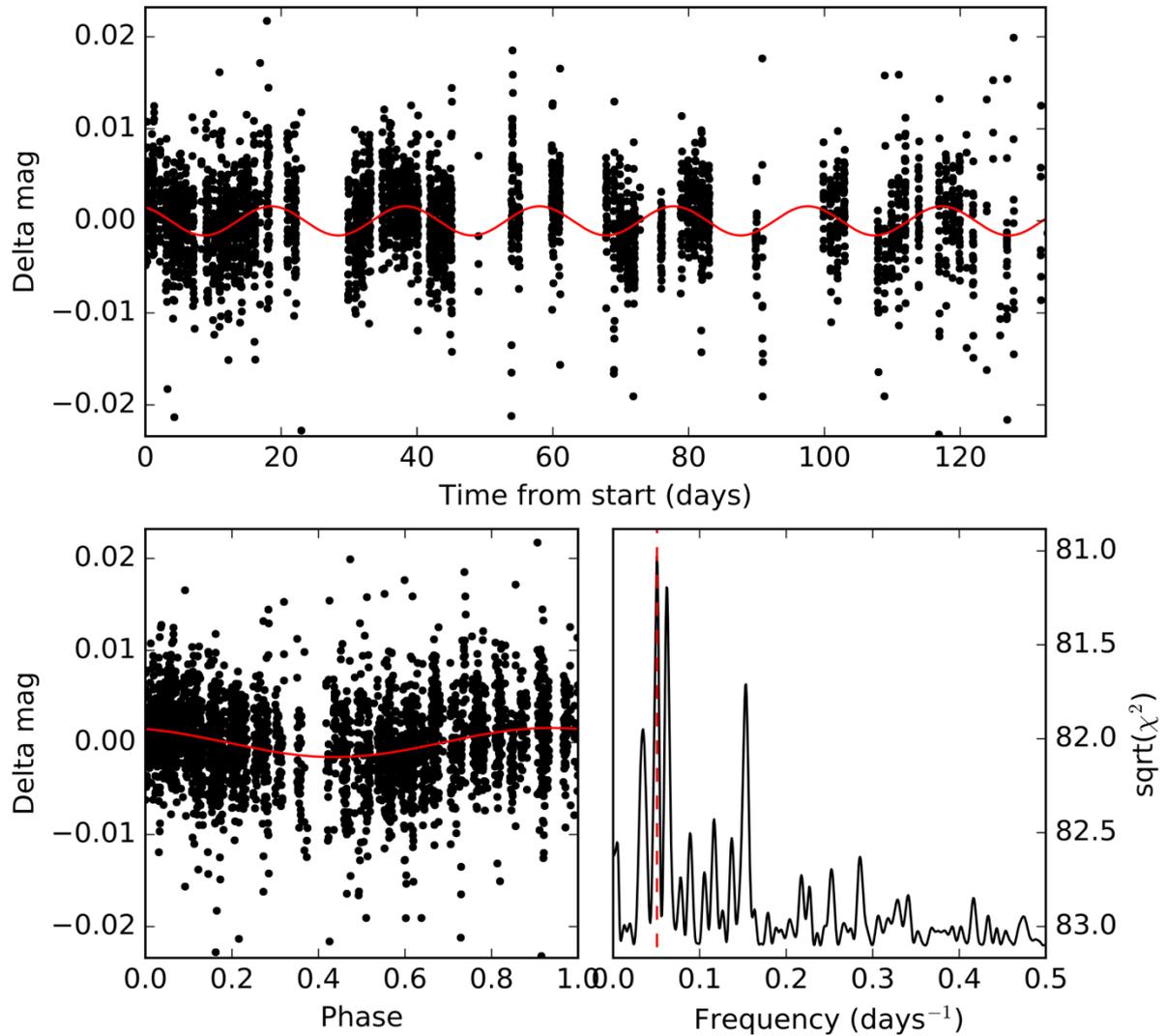

**Extended Data Figure 3 | Ground-based photometry for WASP-107 from MEarth.** We performed a Lomb-Scargle periodogram search and found a best-fit period of 19.7±0.9 days, with a relative amplitude of ~0.00150 mag.

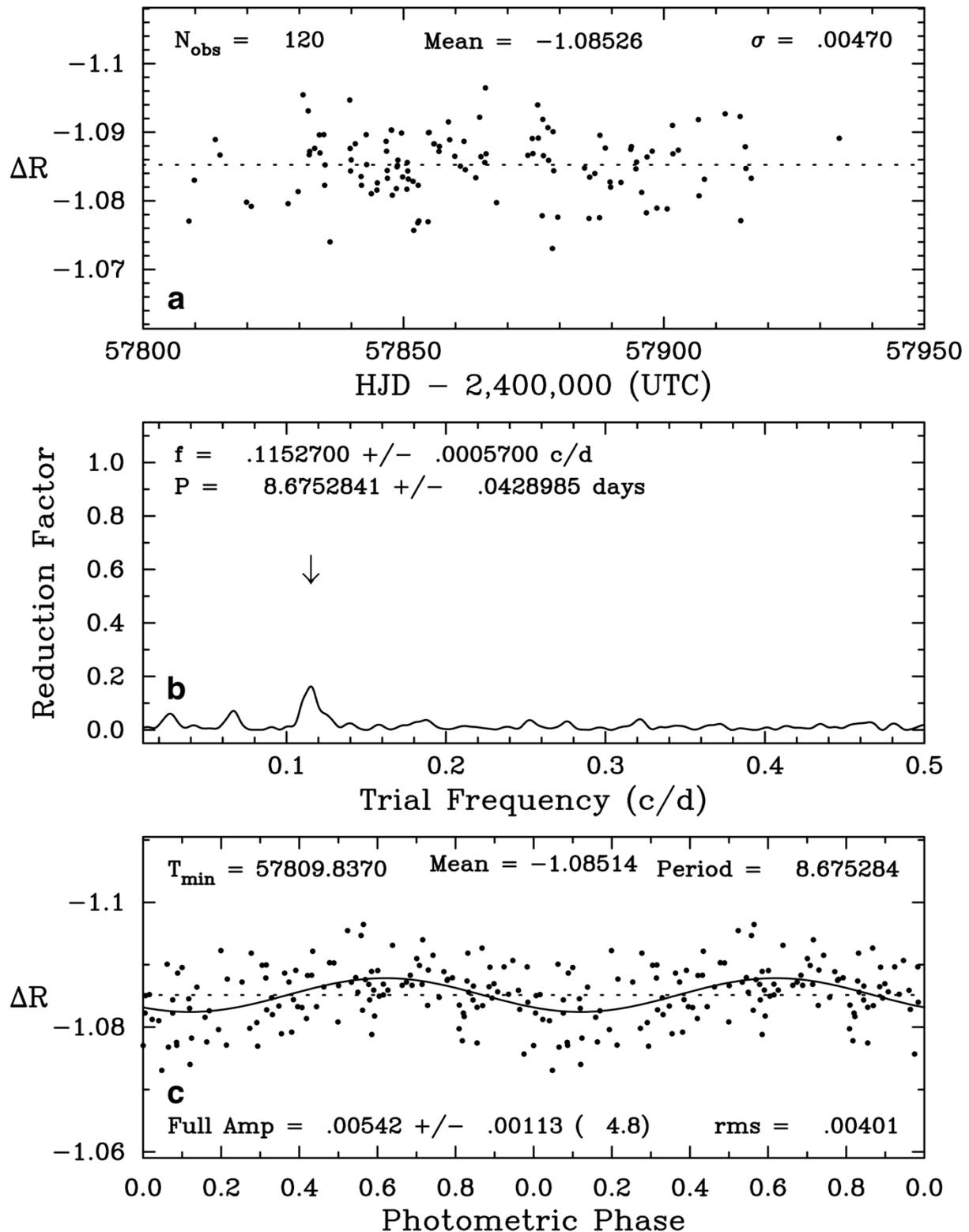

**Extended Data Figure 4 | Ground-based photometry for WASP-107b from AIT.** (a) The nightly photometric observations of WASP-107 in the Cousins R band acquired with the Tennessee State University C14 automated imaging telescope at Fairborn Observatory during the 2017 observing season. (b) The frequency spectrum of the 2017 observations shows low-amplitude variability with a period of 8.675 days. (c) The data phased to the 8.675-day period, has a peak-to-peak amplitude of just 0.005 mag.

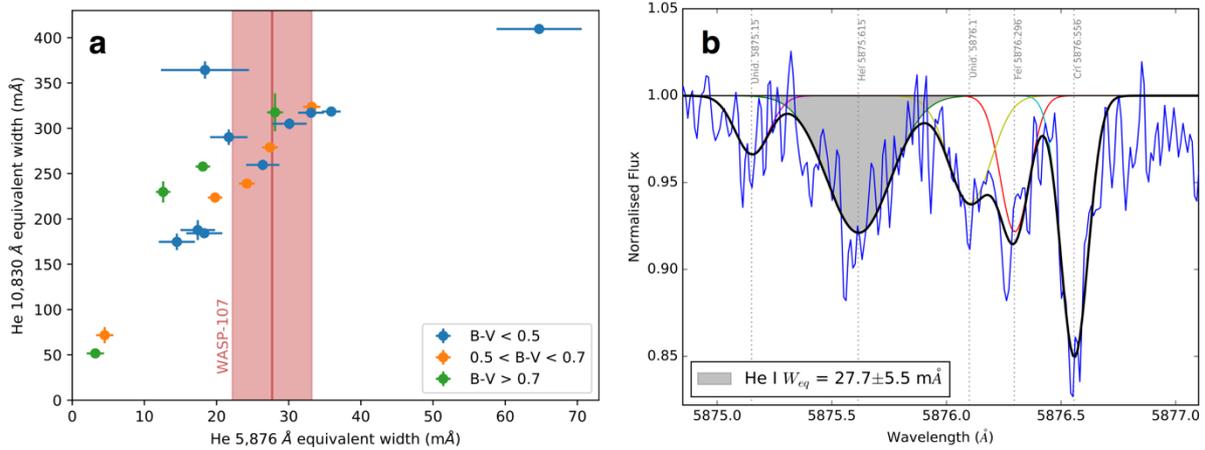

**Extended Data Figure 5 | Equivalent widths of helium 5,876 Å and 10,830 Å lines.** (a) Measurements for 30 stars of different colour indices, from a previous work[63]. These two helium lines are expected to form in the same regions of stellar atmospheres and their equivalent widths are clearly correlated. Our 5,876 Å measurement for WASP-107 is plotted as a red line. Red shaded region shows the 1σ error. Equivalent width measurement and 1σ error of the 5,876 Å line for WASP-107 (B-V > 0.7) from HARPS spectra is shown as red shaded region. (b) Co-added spectra from HARPS radial velocity campaign for WASP-107 around the 5,876 Å line of metastable helium. Lines fit with Gaussian profiles.

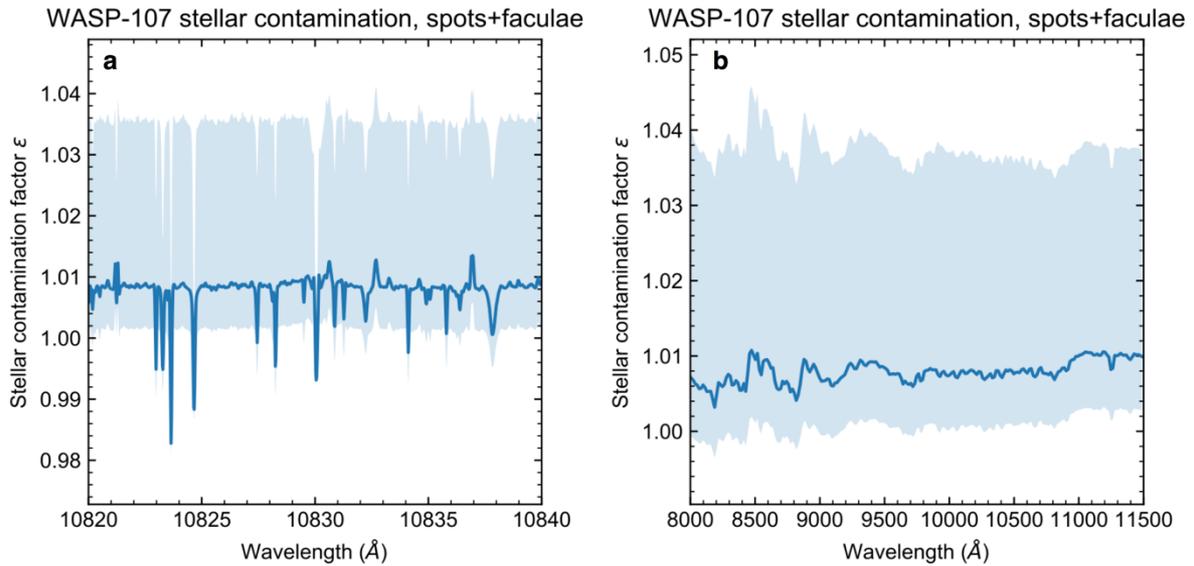

**Extended Data Figure 6 | The effects of an inhomogeneous photosphere on the transmission spectrum of WASP-107b.** Lines show the stellar contamination produced by unocculted spots and faculae. Shaded regions indicate the 1σ uncertainty on the stellar contamination due to the uncertainty on spot and faculae covering fractions. (a) The region around the 10,830 Å (air wavelength) helium triplet at the resolution of the PHOENIX spectra (R=500,000). (b) The full G102 wavelength range in 15 Å bins.

| Parameter | Value |
|---|---|
| $R_P/R_S$ | 0.142988±0.00012 |
| $t_0$ (BJD$_{UTC}$) | 2,457,904.7295±0.0002 |
| $c_0$ | 1.00004±2e-5 |
| $c_1$ | -0.0018±0.0002 |
| $s_{pop}$ | 62±17 |
| $f_{pop}$ | 42±6 |
| $\delta s$ | -2±10 |
| $\delta f$ | 65±4 |
| $\beta$ | 1.73±0.15 |
| $P$ | 5.72147 [a] |
| $i(°)$ | 89.7 [a] |
| $a/R_S$ | 18.164 [a] |
| $e$ (assumed) | 0 |

**Extended Data Table 1 | Fitted parameters from the G102 white light curve.** Errors quoted encompass 68% of the MCMC samples after burn-in. (a) Parameters fixed from Dai & Winn (2017).

| Wavelength (Å) | Transit depth (%) | Error (%) | RMS (PPM) | RMS/phot. | Correction factor |
|---|---|---|---|---|---|
| 8,769 - 9,063 | 2.0451 | 0.0084 | 326 | 1.178 | 1.007101 |
| 9,063 - 9,356 | 2.0425 | 0.0069 | 276 | 1.077 | 1.006785 |
| 9,356 - 9,650 | 2.0514 | 0.0079 | 285 | 1.184 | 1.006549 |
| 9,650 - 9,943 | 2.0514 | 0.0064 | 252 | 1.083 | 1.006454 |
| 9,943 – 10,237 | 2.0456 | 0.0066 | 264 | 1.167 | 1.006340 |
| 10,237 – 10,530 | 2.0448 | 0.0058 | 241 | 1.080 | 1.006303 |
| 10,530 – 10,775 | 2.0431 | 0.0065 | 245 | 1.048 | 1.006162 |
| 10,873 – 11,142 | 2.0461 | 0.007 | 269 | 1.152 | 1.006123 |
| 11,142 – 11,386 | 2.0509 | 0.0069 | 298 | 1.198 | 1.005945 |
| | | | | | |
| 10,579 - 10,677 | 2.0634 | 0.0091 | 344 | 0.989 | 1.00596 |
| 10,604 - 10,701 | 2.0500 | 0.0088 | 381 | 1.102 | 1.005923 |
| 10,628 - 10,726 | 2.0604 | 0.0089 | 366 | 1.061 | 1.006214 |
| 10,652 - 10,750 | 2.0571 | 0.0075 | 336 | 0.976 | 1.006167 |
| 10,677 - 10,775 | 2.0563 | 0.0082 | 360 | 1.043 | 1.006131 |
| 10,701 - 10,799 | 2.0643 | 0.0103 | 395 | 1.143 | 1.006046 |
| 10,726 - 10,824 | 2.0830 | 0.0094 | 354 | 1.023 | 1.005985 |
| 10,750 - 10,848 | 2.0964 | 0.0102 | 415 | 1.198 | 1.005928 |
| 10,775 - 10,873 | 2.1048 | 0.0097 | 391 | 1.126 | 1.005923 |
| 10,799 - 10,897 | 2.0998 | 0.0084 | 387 | 1.117 | 1.005948 |
| 10,824 - 10,922 | 2.0870 | 0.0091 | 390 | 1.128 | 1.005949 |
| 10,848 - 10,946 | 2.0585 | 0.0095 | 409 | 1.183 | 1.006008 |
| 10,873 - 10,970 | 2.0546 | 0.0104 | 385 | 1.111 | 1.005982 |
| 10,897 - 10,995 | 2.0634 | 0.0108 | 423 | 1.220 | 1.005973 |
| 10,922 - 11,019 | 2.0642 | 0.0098 | 377 | 1.087 | 1.005967 |
| 10,946 - 11,044 | 2.0543 | 0.0093 | 363 | 1.046 | 1.005935 |
| 10,970 - 11,068 | 2.0502 | 0.0101 | 375 | 1.084 | 1.005962 |
| 10,995 - 11,093 | 2.0584 | 0.0103 | 373 | 1.082 | 1.005918 |
| 11,019 - 11,117 | 2.0564 | 0.0098 | 385 | 1.117 | 1.005897 |
| 11,044 - 11,142 | 2.0631 | 0.0105 | 414 | 1.197 | 1.005891 |
| **Modified Kreidberg et al. (2017) results:** | | | | | |
| 11,210 – 11,450 | 2.0723 | 0.0059 | | | 1.003979 |
| 11,450 – 11,710 | 2.0814 | 0.0055 | | | 1.003919 |
| 11,710 – 11,960 | 2.0585 | 0.0056 | | | 1.003918 |
| 11,960 – 12,220 | 2.0577 | 0.0054 | | | 1.003848 |
| 12,220 – 12,480 | 2.0535 | 0.0059 | | | 1.003892 |
| 12,480 – 12,720 | 2.0572 | 0.0050 | | | 1.003897 |
| 12,720 – 12,980 | 2.0699 | 0.0062 | | | 1.003830 |
| 12,980 – 13,230 | 2.0818 | 0.0050 | | | 1.003805 |
| 13,230 – 13,490 | 2.0742 | 0.0057 | | | 1.003983 |
| 13,490 – 13,740 | 2.0943 | 0.0048 | | | 1.004081 |
| 13,740 – 14,010 | 2.0878 | 0.0048 | | | 1.004059 |
| 14,010 – 14,250 | 2.0974 | 0.0052 | | | 1.004110 |
| 14,250 – 14,520 | 2.0907 | 0.0062 | | | 1.004126 |
| 14,520 – 14,760 | 2.0777 | 0.0051 | | | 1.004136 |
| 14,760 – 15,020 | 2.0767 | 0.0069 | | | 1.004107 |
| 15,020 – 15,280 | 2.0762 | 0.0067 | | | 1.004020 |
| 15,280 – 15,520 | 2.0593 | 0.0060 | | | 1.004116 |
| 15,520 – 15,790 | 2.0562 | 0.0064 | | | 1.004007 |
| 15,790 – 16,030 | 2.0581 | 0.0056 | | | 1.003941 |
| 16,030 – 16,290 | 2.0595 | 0.0065 | | | 1.003969 |

**Extended Data Table 2 | All results from transit light curve fits.** Modified results from a previous study[18] are included. RMS is the root mean squared of the model residuals in parts per million (PPM); the second-to-last column is the RMS divided by the expected photon noise; the last column is the correction factor we applied to account for stellar variability.

| Parameter | Limits from MCMC |
|---|---|
| Temperature (K) | $650^{+120}_{-80}$ |
| $R_p/R_s$ at 1mbar | $0.914^{+0.010}_{-0.014}$ |
| VMR $\log_{10}(H_2O)$ | $-1.7^{+0.3}_{-0.6}$ |
| VMR $\log_{10}(CO_2)$ | <-10 |
| VMR $\log_{10}(CO)$ | <-11 |
| VMR $\log_{10}(CH_4)$ | <-10 |
| VMR $\log_{10}(NH_3)$ | <-10 |
| VMR $\log_{10}(H_2S)$ | <-11 |
| VMR $\log_{10}(HCN)$ | <-11 |
| VMR $\log_{10}(C_2H_2)$ | <-10 |

**Extended Data Table 3 | Results from ATMO retrieval code for the lower atmosphere.** VMR stands for volume mixing ratio. Uncertainties for temperature, $R_p/R_s$ and VMR $H_2O$ encompass 68% of the MCMC samples after burn-in. Upper limits are from 1σ MCMC errors.